\documentclass{article} % For LaTeX2e
\usepackage{GEM_workshop_2024,times}

\usepackage{amsmath,amsfonts,bm}

\def\eqref#1{equation~\ref{#1}}
\def\1{\bm{1}}

\DeclareMathAlphabet{\mathsfit}{\encodingdefault}{\sfdefault}{m}{sl}
\SetMathAlphabet{\mathsfit}{bold}{\encodingdefault}{\sfdefault}{bx}{n}

\usepackage{hyperref}
\usepackage{url}
\usepackage{enumitem}
\usepackage[pdftex]{graphicx}
\usepackage{adjustbox}
\usepackage{multirow}
\usepackage{booktabs}
\usepackage{fixltx2e}

\title{Protein binding affinity prediction under \\
multiple substitutions applying eGNNs on Residue and Atomic graphs combined with \\
Language model information: eGRAL}

\author{Arturo Fiorellini-Bernardis$^{\dagger\ddagger}$\thanks{Equal Contributions} \And Sebastien Boyer$^{\dagger\ddagger*}$ \And Christoph Brunken\thanks{InstaDeep Ltd, 5 Merchant Square, London, W2 1AY}  \And Bakary Diallo$^\dagger$ \And Karim Beguir$^\dagger$ \And Nicolas Lopez-Carranza$^\dagger$ \And Oliver Bent$^{\dagger}$\thanks{Corresponding Authors: \{a.fiorellini, s.boyer, o.bent\}@instadeep.com}  \\
}

\iclrfinalcopy % Uncomment for camera-ready version, but NOT for submission.
\begin{document}

\maketitle

\begin{abstract}
Protein-protein interactions (PPIs) play a crucial role in numerous biological processes. Developing methods that predict binding affinity changes under substitution mutations is fundamental for modelling and re-engineering biological systems. Deep learning is increasingly recognized as a powerful tool capable of bridging the gap between in-silico predictions and in-vitro observations. With this contribution, we propose eGRAL, a novel SE(3) equivariant graph neural network (eGNN) architecture designed for predicting binding affinity changes from multiple amino acid substitutions in protein complexes. eGRAL leverages residue, atomic and evolutionary scales, thanks to features extracted from protein large language models. To address the limited availability of large-scale affinity assays with structural information, we generate a simulated dataset comprising approximately 500,000 data points. Our model is pre-trained on this dataset, then fine-tuned and tested on experimental data. 

\end{abstract}

\section{Introduction}
Protein-protein interactions (PPIs) are pivotal in many biological processes, such as immune system regulation, cell metabolism, signal transduction and DNA replication~\citep{interactome}. While experimentally measuring these interactions via in-vitro testing is key for advancing modern therapeutics such as cancer therapies \citep{cancerppi}, vaccine design \citep{vaccineppi} and understanding viral infections \citep{viralppi}, these experiments remain a costly and low-throughput process. To circumvent issues of cost and scalability, scientists employ computational models to study the impact of mutations on protein complexes. Existing models generate scores that measure the discrepancy in Gibbs free energy $\Delta G$ between the bound and unbound states of the complex, as well as the variation between mutant (MUT) and wild-type (WT) states $\Delta\Delta G = \Delta G_{MUT}-\Delta G_{WT}$. Most of these models utilize molecular dynamics simulations or simulation-based scoring methods, either as an energy function-based evaluator or integrated within a simulation framework \citep{CCharPPI}. Nevertheless, recent progress in classical machine learning and deep learning techniques has led to the creation of state-of-the-art models in the field of PPI prediction, offering enhanced speed and accuracy \citep{review}. 

Classical machine learning techniques still hold relevance for protein property prediction problems, an example being mmCSM-PPI \citep{mmCSM-PPI}, an Extra Trees based model, leveraging handcrafted features. Since tree-based models cannot handle multiple mutations, mmCSM-PPI addresses this challenge by averaging its handcrafted features over the number of point substitutions. Additionally, the literature offers hybrid approaches that combine deep learning techniques for feature extraction and classical machine learning (mainly tree-based models) to output binding scores \citep{geoppi}, \citep{toptree}. The feature extraction process is either directly integrated into the training pipeline or trained separately in an unsupervised manner. Yet another approach consists in designing deep learning models to predict changes in affinity of PPIs affinity. However, a significant limitation in the majority of these models is their ability to make predictions exclusively for single-point mutations. A notable exception to this trend is NERE \citep{NERE}, which stands out for its unique approach. Trained in a fully unsupervised manner, NERE predicts absolute binding affinity using the structure of the mutated, designed or desired protein complex directly, without explicitly modelling mutations.

The dependency of a model on the existence of a structure represents a limitation: in the case of predicting the effects of mutations, the resulting protein spatial conformation is altered, and since the availability of data from mutated structures is scarce one needs to rely on other tools, such as simulations through classical physics or deep learning based methods. For instance, NERE predicts the absolute binding energy by directly using the complex structure, and in case one wanted to use NERE to evaluate the binding affinity of a structure that has not been experimentally validated, this would likely need to be generated computationally. On the same note, GeoPPI \citep{geoppi} encounters limitations emerging from the fact that it relies on embeddings of both MUT and WT amino acids, which the authors map with a graph neural network (GNN). The solution we propose is based on the adoption of a previously described model, which naturally handles multiple amino acid substitutions as well as the resulting perturbations to the structure thanks to its architecture based on GNNs \citep{egnn}. In this paper, we extend the model to not only combine information coming from atomic and residue scales, but also from the evolutionary scale by adding ESM2-generated \citep{esm} amino acid features: ESM-generated embeddings have demonstrated their usefulness when used in models predicting protein biophysical properties, as shown in \citep{NERE} and \citep{mutateeverything}. Notably, eGRAL utilizes only the WT structure to determine $\Delta\Delta G$, as the MUT features are directly encoded in the WT graph. 

To address the challenge of limited training data, we constructed our own corpus of 519404 $\Delta\Delta G$ scores from single point mutations via Rosetta \citep{das2008macromolecular} simulation on top of SKEMPI\textsubscript{v2} \citep{jankauskaite2019skempi}, as we believe such a Rosetta based score to be most physically accurate for single point mutations. The model is first pre-trained on the \textit{simulated} Rosetta dataset and then fine-tuned on single \textit{and} multiple mutations from the \textit{experimental} SKEMPI\textsubscript{v2} dataset.

\section{Methods}
%\subsection{Cleaned Skempi v2 }
\subsection{eGRAL details}
The architecture of eGRAL follows that proposed by \citep{egnn}. The model consists of two eGNNs: the first eGNN generates embeddings for each residue's atomic environment (AE) and is trained on atomic graphs in a self-supervised way, the second eGNN is trained on residue graphs constructed from PPI binding affinity datasets and scores mutational effects. The contribution of this work is focused on the development of the scorer eGNN, and extends the previous procedure to include features both from a protein language model and to account for specific partner interactions.

\begin{figure}[h]
\centering
\includegraphics[width=0.75\linewidth]{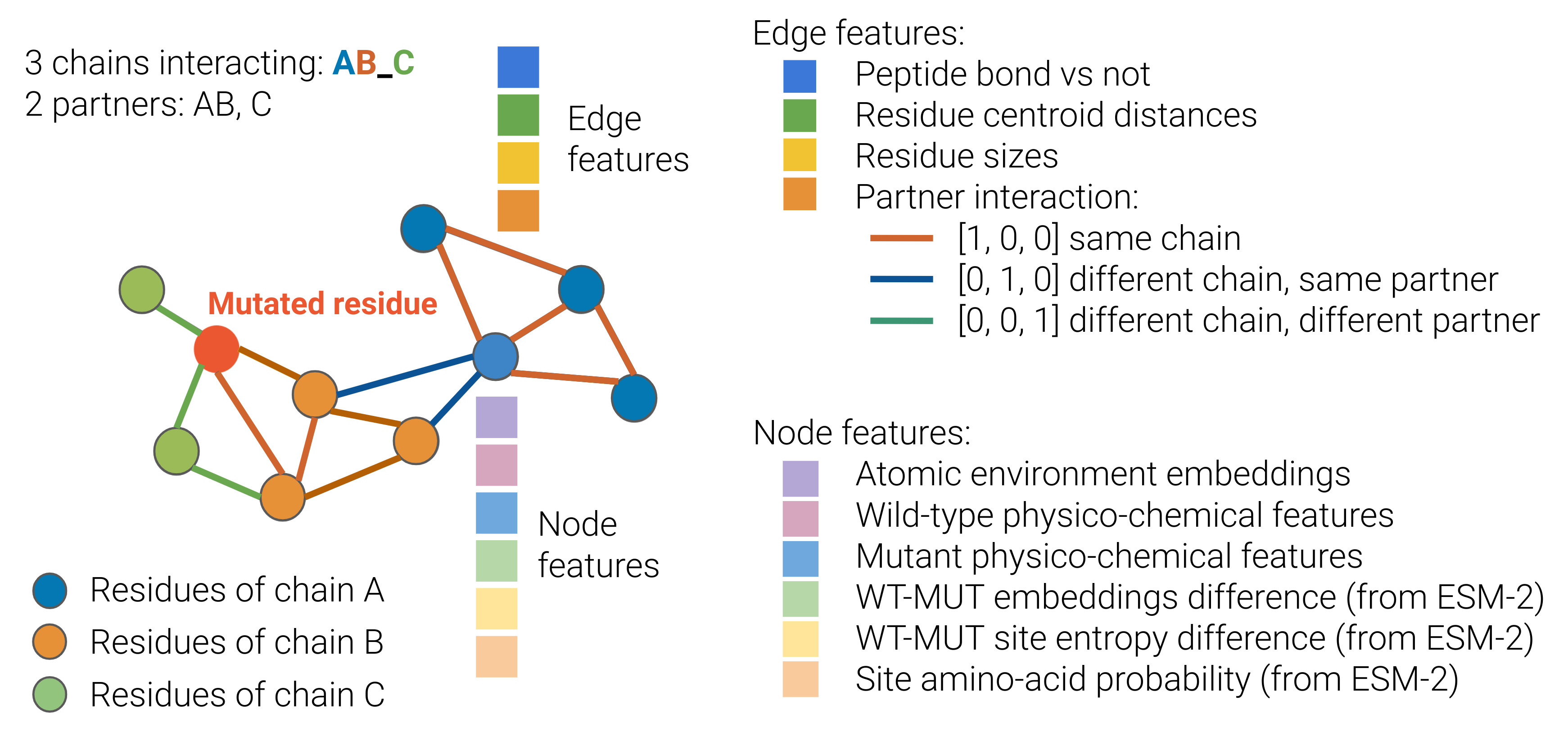}
\caption{Schematic of the residue graphs. Partners AB and C refer to the individual proteins (or sub-units) that interact to form the complex. Each node describes an amino acid represented by multiple features, and edges are drawn between nodes within 9\,\AA. The nodes can include ESM2-generated features while the edges include information on how the partners interact.}
\label{fig:residue_graph}
\centering
\end{figure}

Residue graphs are constructed starting from the MUT residue(s) and drawing edges between residues within a threshold distance of 9\,\AA, Fig. \ref{fig:residue_graph}. The graphs can include N-hop neighbors around the mutated residues, but the results presented with this contribution refer to a 1-hop neighborhood, which we believe is the best trade-off between computational cost and accuracy; in the case of multiple mutations the resulting graphs may be connected or not. The graphs used to train the scorer eGNN are built to exploit the characteristics of protein complexes. In addition to information on the presence of a peptide bond, residue sizes and the distances between amino-acids, the edge features include 1-hot vectors that indicate whether an edge is drawn between residues belonging to the same chain, to different chains in the same partner, or to chains in different partners. Furthermore, node features can include information extracted with protein language models, here ESM2 \citep{esm}: for each amino-acid node, we include the difference between the embeddings of the WT and MUT residues, the difference between the site entropy of the WT and MUT residues, and the site amino-acid probability (when the position is not mutated WT and MUT are the same residue). In particular, the addition of ESM2 generated information increases roughly by a factor 20 the size of the node features: for this reason, we choose the smallest model (esm2\_t6\_8M\_UR50D with 8M parameters) to leverage the higher variance of its predictions and reduce overfitting.

The atomic embedder eGNN is trained following \citep{egnn} but on PDBs processed with a different procedure, explained in Appendix \ref{app:pdb_clean_up}. The multi-layer perceptrons (MLPs) implemented in the scorer eGNN are trained with dropout activated, and the last layer of each MLP is not activated. For further details about the architecture and training hyper-parameters, refer to Appendix \ref{app:hparams}.

\subsection{Data}\label{subs:data}
The PPI binding affinity dataset on which we aim to test eGRAL is SKEMPI\textsubscript{v2}, \citep{jankauskaite2019skempi}. However, the size of the dataset is not sufficient for training the proposed model, thus we resort to a simulated dataset consisting of a library of 519406 protein and variant structures, with single point mutations and scored binding energy changes $\Delta\Delta G$: ROSETTA\textsubscript{sim}. ROSETTA\textsubscript{sim} is constructed with the same PDBs of a cleaned version of SKEMPI\textsubscript{v2}, where all the entries with invalid/ambiguous affinities, ambiguous mutations, non peer-reviewed data, more than one experimental method reported, and PDB IDs with less than 10 data points are deleted: SKEMPI\textsubscript{cl}. Both ROSETTA\textsubscript{sim} and SKEMPI\textsubscript{cl} are divided following the same training, validation, test split (listed in Appendix \ref{app:data_split}): the splits are generated randomly per PDB ID, without any other consideration for any kind of similarity metric between the PDBs themselves. This is done to  ensure that information does not leak through presence of the same structures between splits. For each dataset, the subsets will have subscripts indicating the purpose (\emph{e.g.} ROSETTA\textsubscript{sim,train}, ROSETTA\textsubscript{sim,valid}, ROSETTA\textsubscript{sim,test}). Finally, we generate a test set with experimentally measured binding affinities from \citep{DESAI} and \citep{Starr2022}, consisting in 700 points and up to 7 mutations (referred to as RBD\textsubscript{test}). For a detailed explanation of the derivation of these datasets, refer to Appendix \ref{app:dataset_generation}. 

\subsection{Pre-training and fine-tuning}
To assess the performance and flexibility of the added ESM features, we pre-train, fine-tune and test two different models: one without ESM features (referred to as eGRAL-noESM), one including ESM features (referred to as eGRAL-ESM). As opposed to \citep{egnn}, where the scores were transformed via a Fermi-Dirac function, we train the two models directly on ROSETTA\textsubscript{sim} $\Delta\Delta G$ scores with an AdamW optimizer and using a simple L2 loss. The model checkpoint is chosen according to the lowest L2 loss on the validation set. Model hyper-parameters are listed in Table \ref{tab:h-params}. After the pre-training phase, we use LoRa \citep{lora} to fine-tune the two models with experimental data from SKEMPI\textsubscript{cl}. During fine-tuning we also use a L2 loss and AdamW optimizer, and choose the best model according to the lowest validation loss. Fine-tuning model hyper-parameters are the same used during pre-training, and can be found in Table \ref{tab:h-params}.\\

\section{Results and Discussion}\label{sec:results}
The results section is divided between the pre-training and fine-tuning phases, in order to highlight the different behaviour of the model in the two stages. The models eGRAL-noESM and eGRAL-ESM are trained to minimize the L2 loss, and the best model is chosen to have the lowest loss on the validation set. The results are shown during both phases and for both models in Tables \ref{tab:table_pre_trained} and \ref{tab:table_fine_tuned}; the Spearmank rank correlation coefficient is also reported. During pre-training, the models are trained, validated and tested on ROSETTA\textsubscript{sim}, and can only be considered as an emulator of the configured Rosetta scorer used in this work. It is thus worth noting that the intersection between ROSETTA\textsubscript{sim} and SKEMPI\textsubscript{cl} has a Spearman correlation coefficient of only 0.37 (Appendix: \ref{app:skempi_rosetta_intersection}).

\subsection{Pre-trained model}
The metrics used to monitor the pre-training of the models are shown in Table \ref{tab:table_pre_trained}. Pre-training is done on the simulated dataset as explained in Subsection \ref{subs:data} and detailed in Appendix \ref{app:dataset_generation}. Pre-trained models are tested on three different datasets: ROSETTA\textsubscript{sim,test}, SKEMPI\textsubscript{cl,test} and RBD\textsubscript{test}. This aims at understanding the generalization over multiple degrees of out of distribution data: as ROSETTA\textsubscript{sim,test} also comes from simulated data, we assume its distribution will be closer to the training and validation sets than SKEMPI\textsubscript{cl,test} or RBD\textsubscript{test}, which instead contains experimental values. 

%eGRAL-noESM model achieves a Pearson correlation $\rho_p$ of 0.43, 0.34 and 0.18 on the three test sets, respectively. eGRAL-ESM achieves a Pearson correlation $\rho$ of 0.40, 0.46 and 0.17 on the aforementioned three test sets, respectively. 
Due to its high degree of expressiveness, eGRAL-ESM overfits the training set (ROSETTA\textsubscript{sim,train} $\rho_p$: 0.69, ROSETTA\textsubscript{sim,valid} $\rho_p$: 0.50) and does not perform significantly better than eGRAL-noESM model on ROSETTA\textsubscript{sim,test} (eGRAL-noESM $\rho_p$: 0.43, eGRAL-ESM $\rho_p$: 0.40) but does on SKEMPI\textsubscript{cl,test} (eGRAL-noESM $\rho_p$: 0.34, eGRAL-ESM $\rho_p$: 0.46). Both models show rather poor performance over RBD\textsubscript{test}. Appendix \ref{app:rosetta_test} reports figures with a more granular analysis of the performance of the two pre-trained models over ROSETTA\textsubscript{sim,test}. 

\begin{table*}[h]
\caption{Evaluation metric summary for the pre-trained model. RMSE is expressed in kcal/mol, $\rho_p$ is the Pearson correlation coefficient and $\rho_s$ is the Spearman rank correlation coefficient.}
\label{tab:table_pre_trained}
\vskip 0in
\begin{center}
\setlength{\tabcolsep}{3pt} 
\begin{scriptsize}
\begin{sc}
\begin{tabular}{l|ccc||ccc||ccc|ccc|ccc}
\toprule
\multirow{2}{*}{} & \multicolumn{3}{|c}{ROSETTA\textsubscript{sim,train}} & \multicolumn{3}{|c}{ROSETTA\textsubscript{sim,valid}} & \multicolumn{3}{|c}{ROSETTA\textsubscript{sim,test}} & \multicolumn{3}{|c}{SKEMPI\textsubscript{cl,test}} & \multicolumn{3}{|c}{RBD\textsubscript{test}}\\
\cline{2-16}
 & RMSE & $\rho_p$ & $\rho_s$ & RMSE & $\rho_p$ & $\rho_s$ & RMSE & $\rho_p$ & $\rho_s$ & RMSE & $\rho_p$ & $\rho_s$ & RMSE & $\rho_p$ & $\rho_s$ \\ 
\midrule
$eGRAL_{no ESM}^{pre-trained}$ & 2.14 & 0.46 & 0.37 & 1.92 & 0.50 & 0.39 & \textbf{2.11} & \textbf{0.43} & \textbf{0.33} & \textbf{1.87} & \textbf{0.34} & \textbf{0.41} & 0.90 & \textbf{0.18} & \textbf{0.14} \\
$eGRAL_{ESM}^{pre-trained}$ & 1.70 & 0.70 & 0.57 & 1.87 & 0.49 & 0.34 & 2.19 & 0.38 & 0.28 & 2.01 & 0.30 & 0.33 & \textbf{0.77} & 0.08 & 0.05 \\
\bottomrule
\end{tabular}
\end{sc}
\end{scriptsize}
\end{center}
\vskip -0.15in
\end{table*}

%%%%%%%%%%%%%%%%%%%%%%%%%%%%%%%%%%%%%%%%%%%%%%%%%%%%
%%%%%%%%%%%%%%%%%%%%%%%%%%%%%%%%%%%%%%%%%%%%%%%%%%%%
\subsection{Fine tuned model}
Following pre-training, the model is fine-tuned with LoRA on multiple mutations and experimental values from SKEMPI\textsubscript{cl}. Results are shown in Table \ref{tab:table_fine_tuned}. With the fine-tuning procedure, both eGRAL-noESM and eGRAL-ESM achieve an increased Pearson correlation coefficient over SKEMPI\textsubscript{cl,test}, that goes from 0.34 to 0.47 for the former, and from 0.46 to 0.57 for the latter. Overall, neither model's performance improves for the RBD\textsubscript{test} dataset. During pre-training, eGRAL-ESM overfits the training set rapidly, which makes the fine-tuning procedure difficult. For this reason, the model does not show the improved performance which may have been expected from the inclusion of the ESM2 features.

\begin{figure}[h]
\centering
\includegraphics[width=0.9\linewidth]{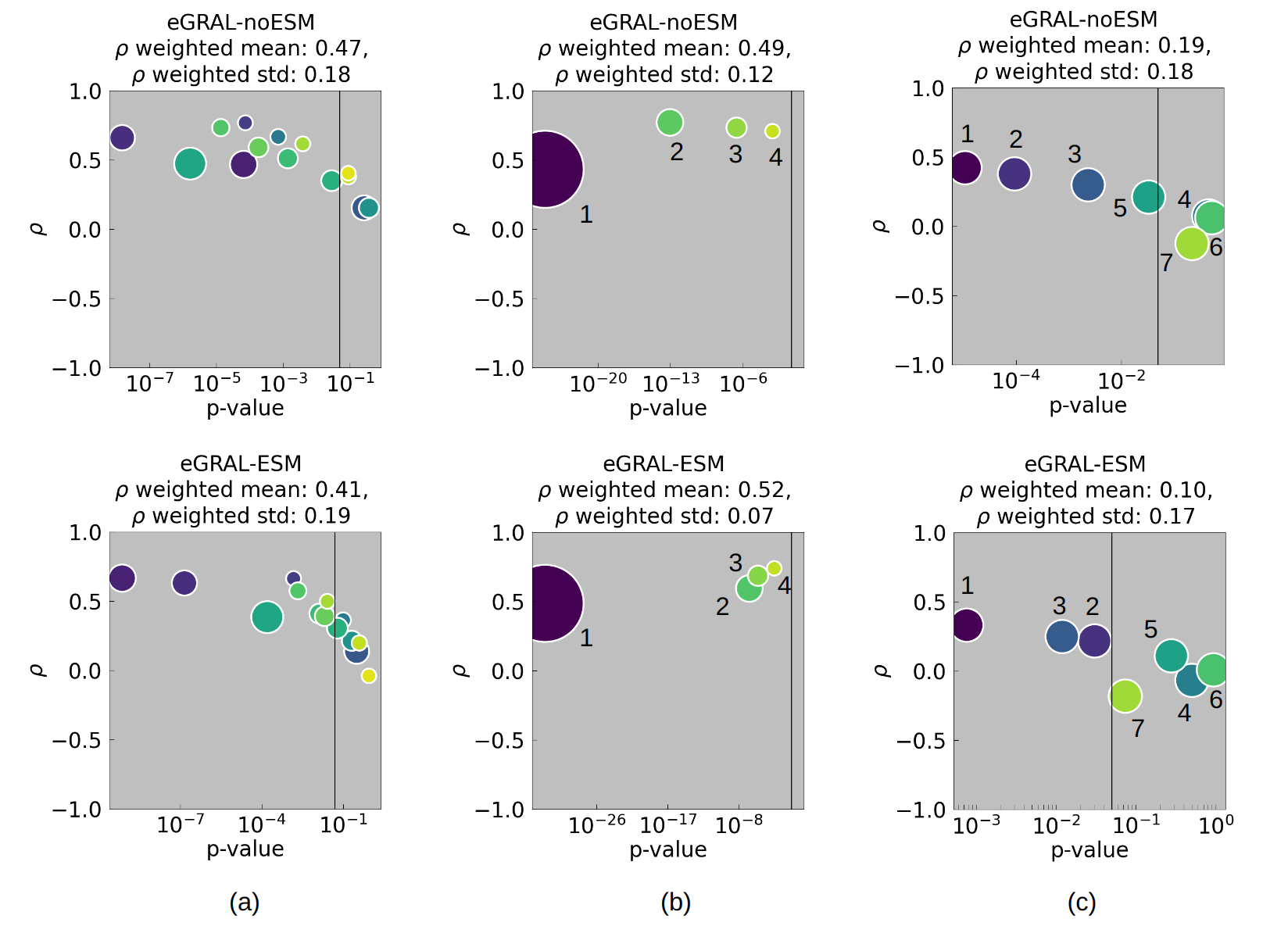}
\caption{Performance of fine-tuned eGRAL-ESM (top row) and eGRAL-noESM (bottom row). Results are broken down as follows; (a) per PDB; (b) per number of mutations for SKEMPI\textsubscript{cl,test}; and (c) per number of mutations for RBD\textsubscript{test}. The Pearson correlation coefficient ($\rho$) is reported. Marker size is proportional to the number of points used to calculate the correlation. The colouring refers to the different PDB IDs or number of mutations. In column (b) and (c), the numbering refers to number of mutations. The vertical line marks a significant p-value of 0.05.}
\label{fig:freq_cav_data}
\centering
\end{figure}

For both models the test sets are then broken down by PDB IDs and number of mutations for SKEMPI\textsubscript{cl,test}, while only per number of mutations for RBD\textsubscript{test}, in Figure~\ref{fig:freq_cav_data}. The predictive power of both models, measured with the Pearson correlation coefficient, does not strongly depend on the identity of the PDB, which shows that the models can generalise to diverse protein complexes (for further details refer to Appendix \ref{app:skempi}). Figure~\ref{fig:freq_cav_data} also shows the average Pearson correlation coefficient weighted by the number of data points per PDB, and seperately per number of mutations. In terms of predictive power conditional to the number of mutations to score, both models have significant Pearson correlation coefficient up to four substitutions on SKEMPI\textsubscript{cl,test}. On the contrary, although Figure~\ref{fig:freq_cav_data} might indicate that both models produce meaningful predictions in case of multiple substitutions over RBD\textsubscript{test}, Figures~\ref{fig:preds_per_mut_rbd_noesm} and \ref{fig:preds_per_nmut_rbd_esm8m} in Appendix \ref{app:rbd_test} show that this is not the case: indeed, the models output meaningful results only in the case of single point mutations.

\renewcommand{\arraystretch}{1.1}
\begin{table*}[h]
\caption{Evaluation metric summary for the fine tuned model. RMSE is expressed in kcal/mol, $\rho_p$ is the Pearson correlation coefficient and $\rho_s$ is the Spearman rank correlation coefficient.}
\label{tab:table_fine_tuned}
\vskip 0.1in
\begin{center}
\setlength{\tabcolsep}{3pt} 
\begin{scriptsize}
\begin{sc}
\begin{tabular}{l|ccc|ccc|ccc|ccc}
\toprule
\multirow{2}{*}{} & \multicolumn{3}{|c}{SKEMPI\textsubscript{cl,train}} & \multicolumn{3}{|c}{SKEMPI\textsubscript{cl,valid}} & \multicolumn{3}{|c}{SKEMPI\textsubscript{cl,test}} & \multicolumn{3}{|c}{RBD\textsubscript{test}}\\
\cline{2-13}
 & RMSE & $\rho_p$ & $\rho_s$ & RMSE & $\rho_p$ & $\rho_s$ & RMSE & $\rho_p$ & $\rho_s$ & RMSE & $\rho_p$ & $\rho_s$\\ 
\midrule
$eGRAL_{no ESM}^{fine-tuned}$ & 1.73 & 0.48 & 0.50 & 1.78 & 0.49 & 0.57 & 1.77 & 0.47 & 0.42 & 1.00 & \textbf{0.22} & \textbf{0.11}\\
$eGRAL_{ESM}^{fine-tuned}$ & 1.70 & 0.47 & 0.52 & 1.72 & 0.47 & 0.55 & \textbf{1.73} & \textbf{0.50} & 0.42 & \textbf{0.78} & 0.17 & 0.07\\
\bottomrule
\end{tabular}
\end{sc}
\end{scriptsize}
\end{center}
\vskip -0.15in
\end{table*}

\subsection{Discussion}
eGRAL and our training procedure show significant predictive power. Indeed, we see for both models improved prediction of experimental scores after the fine-tuning process. Both models generate useful predictions for up to four substitutions over SKEMPI\textsubscript{cl,test}. Regarding RBD\textsubscript{test} the cut is less clear even though we see significant predictive power in the case of single substitutions. Our model outperforms geoPPI (Pearson $\rho$: $-0.18$,  p-value: 0.07) on the single mutation subset from RBD\textsubscript{test}, Fig. \ref{fig:geoPPI}.

The evidence of the advantage of adding ESM features emerges when evaluating the model on SKEMPI\textsubscript{cl,test} after the fine-tuning step, see Table \ref{tab:table_pre_trained} and \ref{tab:table_fine_tuned}. Indeed, while the performance of eGRAL-ESM is worse than eGRAl-noESM after pre-training, the fine-tuning process leads to slightly better results. However, we want to highlight the fact that eGRAL-ESM shows a significantly larger predictive potential thanks to its higher expressivity than eGRAL-noESM: indeed, eGRAL-ESM can achieve Pearson correlation up to ~0.80 over the training set, whereas eGRAL-noESM reaches a plateau at about ~0.50. In our case, this translates into eGRAL-ESM overfitting the dataset quite quickly, but we believe that using a model with ESM features might, in the future, be more appropriate in the context of a larger simulated dataset, probably built around more diverse PDBs and already including multiple amino acid substitutions.

To assess the robustness of our procedure and verify its usefulness compared to the state of the art, we proceed with a variance analysis where we test eGRAL-noESM and eGRAL-ESM pre-trained and fine-tuned with 5 different initialization seeds and 5 different training and validation splits (see Appendix \ref{app:variance}); the models are trained leaving architecture and hyper-parameters fixed as in Appendix \ref{app:hparams}. The 10 resulting models are tested on SKEMPI\textsubscript{cl,test} and the performance assessed with Pearson $\rho_p$ and Spearman rank $\rho_s$ correlation coefficients: the average metrics $\pm$ standard deviation across the seeds result $\rho_p$=0.37$\pm$0.11 and $\rho_s$=0.38$\pm$0.11 for eGRAL-noESM, $\rho_p$=0.47$\pm$0.04 and $\rho_s$=0.40$\pm$0.04 for eGRAL-ESM; the average metrics $\pm$ standard deviation across the splits result $\rho_p$=0.43$\pm$0.05 and $\rho_s$=0.40$\pm$0.05 for eGRAL-noESM, $\rho_p$=0.45$\pm$0.13 and $\rho_s$=0.37$\pm$0.13 for eGRAL-ESM. Even without further hyper-parameter tuning specific to each new seed or split, the lower bound of the performance of our procedure, calculated as mean minus standard deviation, is better or really close to that obtained with our Rosetta scorer config (see Figure \ref{fig:variance_analysis_seeds} and \ref{fig:variance_analysis_splits}, which instead has an RMSE of 2.62 kcal/mol and correlates to SKEMPI\textsubscript{cl,test} with $\rho_p$=0.33 and $\rho_s$=0.37 (Appendix \ref{app:skempi_rosetta_intersection}. Moreover, eGRAL is tested on multiple mutations, whereas the performance of Rosetta is evaluated only for single point mutations, which we feel confident in assuming as an easier scenario, further underlining the robustness of this work. Note that the same analysis is not performed after the pre-training phase since eGRAL would perform at best as a Rosetta emulator at this stage. Finally, we compare the execution speed by scoring all the possible 19 single mutations in a position of a PDB: eGRAL-noESM takes 25s, eGRAL-ESM takes 34s, against an average of 49s (min: 31s, max: 57s) for Rosetta (further details in Appendix \ref{app:time_benchmark}).

Among the limitations of eGRAL, a significant one lies in the fact that it does not work with a MUT structure directly. While the graphs encode information about the size of the residue, see Figure \ref{fig:residue_graph}, allowing the model to understand when the complex is altered, the coordinates of the mutated residues remain the same. We recognize that having a model that extracts information from a mutated structure would be optimal but this does come with costs: experimentally derived structures are of scarce availability, while simulated ones are often difficult to solve as well as noisy; moreover, WT and MUT structure might generate different graphs (e.g. residues within the cut-off could end up outside of it after substitution, and vice-versa), making it cumbersome to compare them. For these reasons, we refrain from using a MUT structure. Another limitation emerges from the data: eGRAL is trained on static coordinates, whereas proteins are inherently dynamic systems. Possible approaches to tackle these issues could be techniques to train via denoising score matching and have the model learn a less static representation of the protein geometry by either adding Gaussian noise \citep{DSMBind}, or rigid transformation noise \citep{NERE} to the PDB atomic coordinates.

\section{Conclusions}
Utilizing a simulated PPI binding score dataset, we pre-trained multiscale eGNN models, an architecture that is flexible to multiple substitutions. We then extended the procedure by introducing a fine-tuning step to have eGRAL learn on limited experimental values. eGRAL shows good predictive power and generalises well to different PDBs as well as multiple mutations. Although adding ESM2 generated features does not significantly improve the performance on the test sets, the model becomes more expressive and we believe evidences potential for further exploration.



\bibliography{iclr2024_conference}
\bibliographystyle{iclr2024_conference}

\appendix
\section{Appendix}
\subsection{Datasets generation}\label{app:dataset_generation}
\textbf{Cleaned SKEMPIv2:}\label{app:skempi_cl} We used the SKEMPI\textsubscript{v2} dataset and passed it through another cleaning step as we found some of its entries unclear. Hence from here on whenever we will talk about the SKEMPI\textsubscript{cl} dataset, we refer to the subset of SKEMPI\textsubscript{v2} dataset for which all entries meeting the following criteria are deleted: invalid/ambiguous affinities, ambiguous mutations, non peer-reviewed data, entries with more than one experimental method reported, all PDBs with mutation count less than ten. Training, validation and test splits are generate by randomly splitting the dataset per PDB ID, without any other consideration for any kind of similarity metric between the PDBs themselves. This is done to  ensures that information does not leak between splits (the same PDB is never shared between different splits). In the following, those splits are referred to as SKEMPI\textsubscript{cl,train}, SKEMPI\textsubscript{cl,valid}, SKEMPI\textsubscript{cl,test}, they are listed in Appendix \ref{app:data_split} and their distribution shown in Fig. \ref{fig:skempi-split}. It is this split that is used to train and finetune the models.

\textbf{Simulated data based on Cleaned SKEMPIv2:}\label{simul_data} A library of protein structure variants was constructed using PDB IDs and their interface definitions, from SKEMPI\textsubscript{cl}. The interface corresponds to two interacting sub-units (partner A and partner B) of the protein chain(s). From their RCSB PDB structures, their interface residues are extracted. An interface residue is defined as any residue on a chain of partner A with an atom within a distance of 8 Å to any other atom on a chain of partner B. These residues are mutated to all possible amino acids, excluding the wild-type. Only single point mutations are considered. This process generates a total library of 541680 variants. The binding energy change $\Delta\Delta G$ upon mutation is then computed using a Rosetta \citep{das2008macromolecular} protocol. All wild type structures are first relaxed using the Rosetta FastRelax protocol. Mutations are then applied to the protein structures, followed by a repacking of the surrounding residues to accommodate the structural changes induced by the mutation. The radius for surrounding residues selection for repacking is set to 7 Å. The structure is then refined post-mutation: a 'backrub' technique using Rosetta BackrubMover module followed by a final minimization step using FastRelax with the 'lbfgs\_armijo\_nonmonotone' algorithm. Parameters for the Monte Carlo steps and temperature, crucial for the Metropolis-Hastings criterion, are set at 500 and 0.4, respectively. The $\Delta\Delta G$ of binding energies is evaluated using Rosetta InterfaceAnalyzerMover module. 'Ref2015' is consistently used for all scoring tasks leading to 519406 variants successfully scored (predicted absolute score less than 12 kcal/mol). The lists of PDB IDs from this dataset used to train, validate and test the model is the same for both the simulated dataset ROSETTA and the experimental dataset SKEMPI\textsubscript{cl}: the split is reported in Appendix \ref{app:data_split}. We use the same split to ensure that there is not data leakage between the pre-training and the fine-tuning phases. Hence, we generate a ROSETTA\textsubscript{sim,train}, ROSETTA\textsubscript{sim,valid} and ROSETTA\textsubscript{sim,test} that we use to pre-train our models.

\textbf{RBD test dataset:} We generate a test set starting from the Desai SARS-CoV-2:ACE2 dataset \citep{DESAI}: the dataset systematically maps the epistatic interactions between the 15 mutations in the receptor binding domain (RBD) of Omicron BA.1 relative to the Wuhan-Hu-1 strain. The dataset include experimental measurements of the ACE2 affinity of all possible combinations of these 15 mutations. From it we generate a subset of up to 7 mutations (100 points per each number of mutations); however, since there are only 15 single point mutation affinity values and we intend to have 100 data points, we merge 85 single substitutions in Wuhan-Hu-1 taken from deep mutational scan measurements from \citep{Starr2022}. The correlation between the overlap in the two datasets is 88\% (Spearman). For each point, the $\Delta K_d$ is calculated using the $K_d$ (dissociation constant) of each mutants referred to the background. For this dataset, the structure used was generated with an Alphafold2 pipeline \citep{AlphaFold2021}. In the following this dataset is referred to as RBD\textsubscript{test}.

\subsection{Rosetta based scorer performance evaluation}\label{app:skempi_rosetta_intersection}
\begin{figure}[h]
\centering
\includegraphics[width=0.35\linewidth]{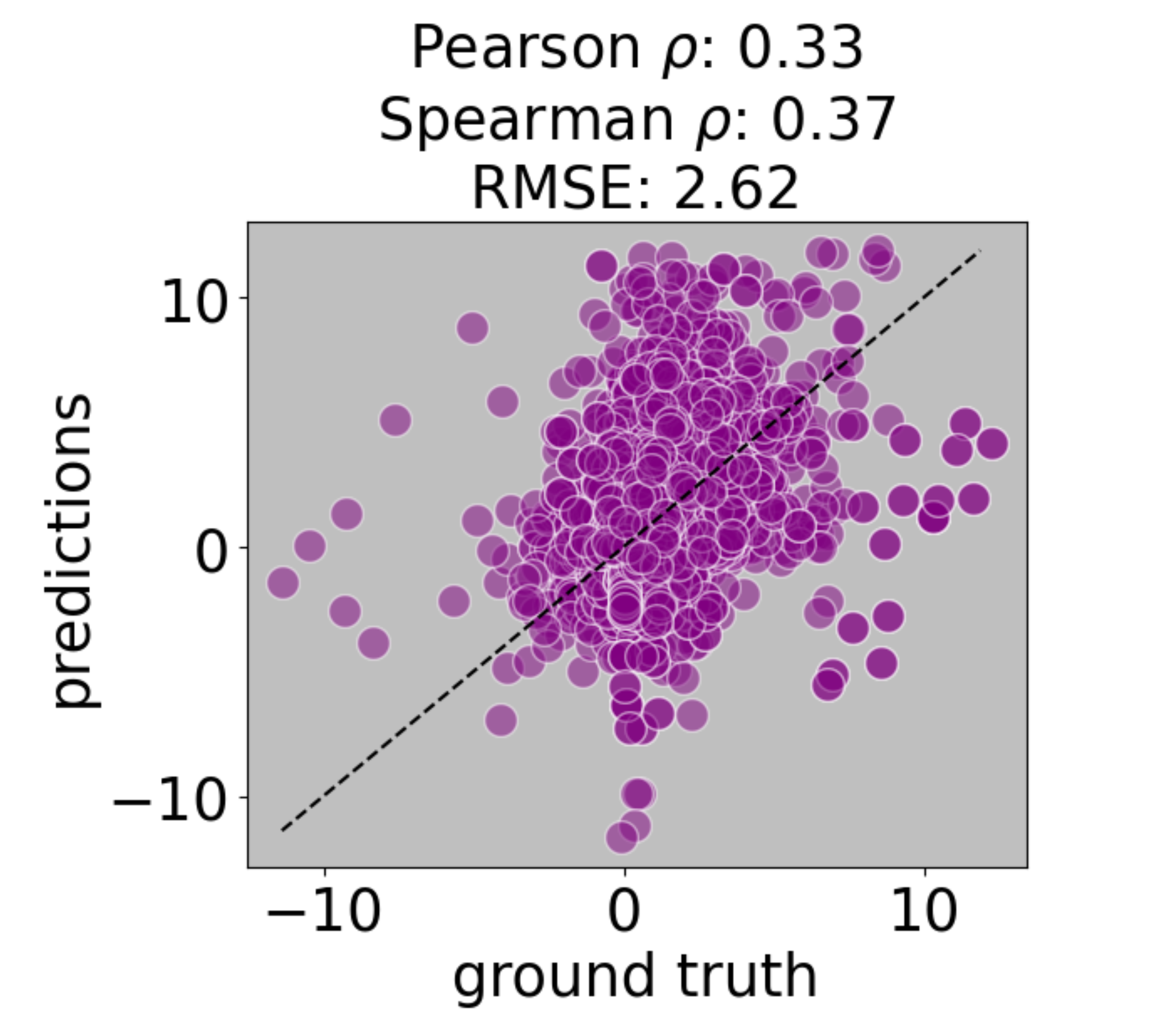}
\caption{$\Delta\Delta G$ predicted by the Rosetta-based scorer against the experimental values in SKEMPI\textsubscript{cl} for the intersection between ROSETTA\textsubscript{sim} and SKEMPI\textsubscript{cl}.}
\label{fig:rosetta_vs_skempi}
\centering
\end{figure}

This appendix shows how well the simulated dataset ROSETTA\textsubscript{sim} correlates to the experimental values in SKEMPI\textsubscript{cl}. The intersection between the two datasets consists in 3741 single point mutations across 229 PDBs. The results are showed in Fig. \ref{fig:rosetta_vs_skempi}. For the intersection, the simulated dataset correlates to SKEMPI\textsubscript{cl} with a RMSE of 2.62 kcal/mol and Pearson and Spearman rank correlation coefficients of 0.33 and 0.37, respectively. We do not score multiple mutations with Rosetta as it is safe to assume that the results would be at best as good, but likely worse, than the single mutation scenario.

\subsection{Data splits}\label{app:data_split}
In this appendix we list in alphabetical order the PDB IDs used for the training, validation and test sets both with the ROSETTA\textsubscript{sim} and SKEMPI\textsubscript{cl} sets:
\begin{itemize}
    \item Training split: \small{1A22, 1A4Y, 1AHW, 1AO7, 1B2S, 1B2U, 1B3S, 1BD2, 1BP3, 1C1Y, 1C4Z, 1CBW, 1CHO, 1CSE, 1CT0, 1CT2, 1DVF, 1EFN, 1F5R, 1FC2, 1FCC, 1FFW, 1FR2, 1FSS, 1FY8, 1GCQ, 1GL0, 1GUA, 1H9D, 1HE8, 1IAR, 1JTG, 1KAC, 1KIQ, 1KIR, 1KTZ, 1LFD, 1LP9, 1M9E, 1MAH, 1MI5, 1MLC, 1N8O, 1N8Z, 1NCA, 1NMB, 1OGA, 1P6A, 1PPF, 1QSE, 1R0R, 1REW, 1S0W, 1S1Q, 1SBB, 1SGN, 1SGP, 1SGY, 1SIB, 1TM3, 1TM4, 1TM5, 1TM7, 1TMG, 1U7F, 1WQJ, 1X1W, 1X1X, 1XGP, 1XGQ, 1XGR, 1XGT, 1XGU, 1Y1K, 1Y3B, 1Y3C, 1Y3D, 1Y48, 1YQV, 1YY9, 1Z7X, 2AJF, 2B0U, 2B10, 2B11, 2B2X, 2B42, 2BNR, 2BTF, 2C5D, 2DSQ, 2DVW, 2E7L, 2G2U, 2G2W, 2GOX, 2HRK, 2I26, 2J0T, 2J12, 2J1K, 2J8U, 2JCC, 2JEL, 2NU0, 2NU1, 2NU2, 2NU4, 2NYY, 2NZ9, 2O3B, 2OI9, 2P5E, 2PCB, 2PCC, 2REX, 2SGP, 2SGQ, 2VIR, 2VIS, 2VLO, 2VLP, 2WPT, 3B4V, 3BN9, 3BT1, 3BTD, 3BTE, 3BTM, 3BTQ, 3BTT, 3BX1, 3D3V, 3D5S, 3EQS, 3F1S, 3G6D, 3H9S, 3HFM, 3HH2, 3LB6, 3M62, 3MZG, 3MZW, 3N06, 3N4I, 3N85, 3NCB, 3NCC, 3NPS, 3NVN, 3NVQ, 3PWP, 3Q3J, 3Q8D, 3QDG, 3QFJ, 3QHY, 3R9A, 3RF3, 3SE3, 3SE4, 3SE8, 3SEK, 3SF4, 3SGB, 3U82, 3UIG, 3WWN, 4CPA, 4E6K, 4EKD, 4G2V, 4GNK, 4GXU, 4HFK, 4HRN, 4HSA, 4J2L, 4JFF, 4JGH, 4K71, 4L0P, 4L3E, 4LRX, 4MYW, 4NM8, 4NZW, 4O27, 4OZG, 4P5T, 4RA0, 4U6H, 4WND, 4X4M, 4Y61, 4YFD, 4YH7, 5C6T, 5CXB, 5CYK, 5E6P, 5K39, 5M2O, 5UFQ}
    \item Validation split: \small{1N80, 1TM1, 2HLE, 2OOB, 2QJA, 2VLQ, 3KBH, 3LZF, 3N0P, 3SZK, 4FZA, 4MNQ, 4OFY, 4UYQ, 5TAR, 5UFE}
    \item Test split: \small{1AK4, 1B41, 1BJ1, 1EMV, 1F47, 1GC1, 1JTD, 1K8R, 1MHP, 1VFB, 2FTL, 2SIC, 3AAA, 3C60, 3L5X, 3NGB, 3SE9, 4FTV, 4P23, 4PWX, 5E9D, 5XCO}
\end{itemize}
Fig. \ref{fig:skempi-split} shows the frequency of data points in SKEMPI\textsubscript{cl} split as above and divided per number of mutation:
\begin{figure}[h]
\centering
\includegraphics[width=1\linewidth]{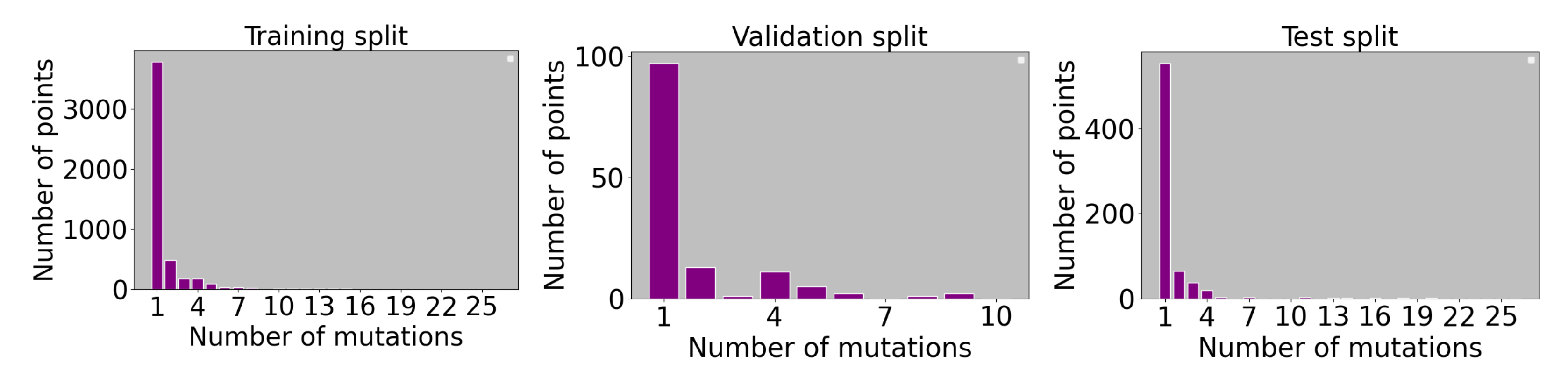}
\caption{Frequency of data points in SKEMPI\textsubscript{cl} per number of mutation in different splits. Namely from left to right: training, validation and test split.}
\label{fig:skempi-split}
\centering
\end{figure}

%%%%%%%%%%%%%%%%%%%%%%%%%%%%%%%%%%%%%%%%%%%%%%%%%%%%%%%%%%%%%%%%%%%%%%%%
%%%%%%%%%%%%%%%%%%%%%%%%%%%%%%%%%%%%%%%%%%%%%%%%%%%%%%%%%%%%%%%%%%%%%%%%
\subsection{ROSETTA\textsubscript{sim} test add-ons}\label{app:rosetta_test}
\begin{figure}[h]
\centering
\includegraphics[width=0.95\linewidth]{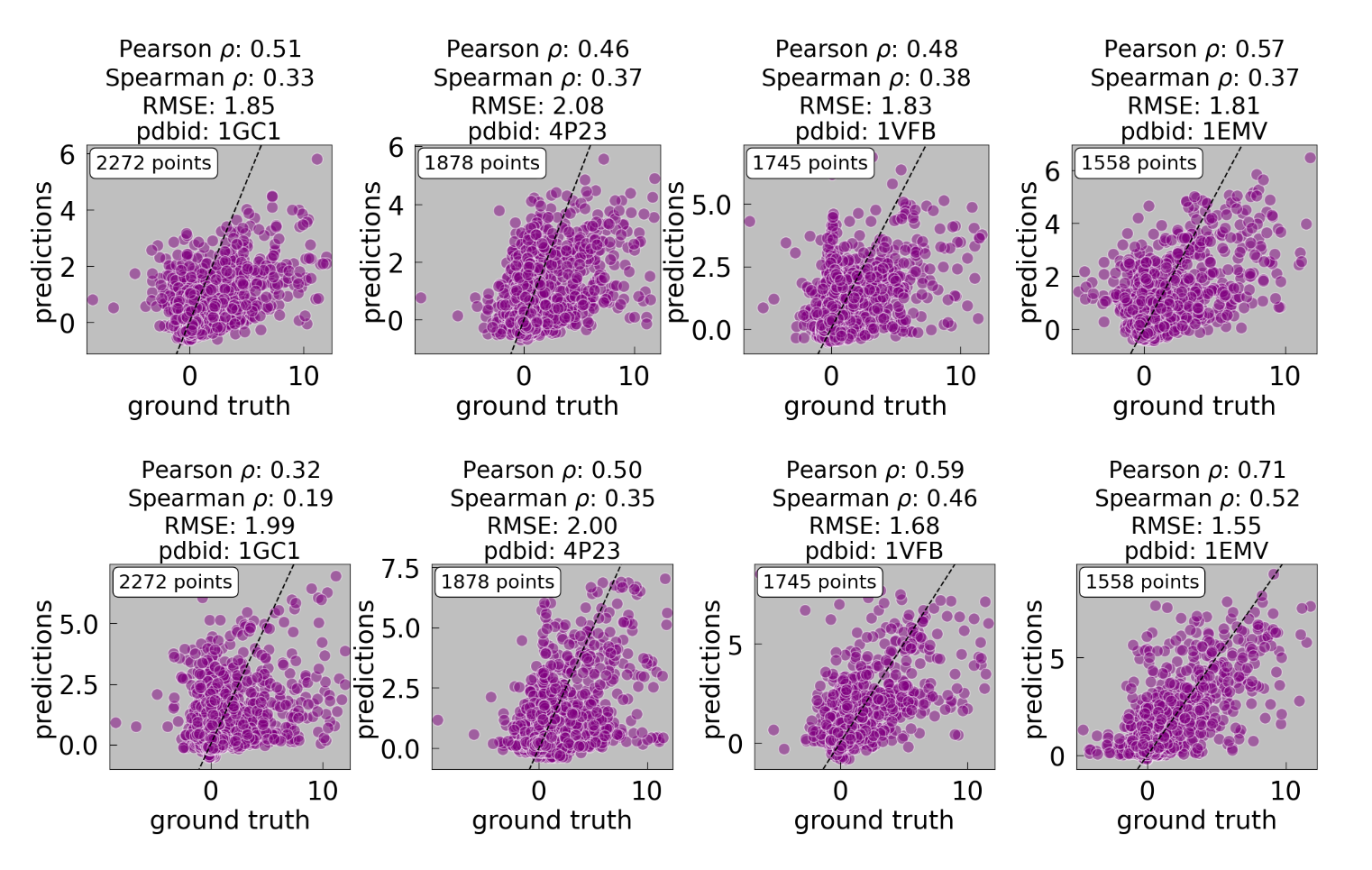}
\caption{Predicted results against ground truth by pre-trained eGRAL-noESM (top row) and eGRAL-ESM (bottom row) on ROSETTA\textsubscript{sim,test}: the predictions are reported for 4 PDB IDs. Scores and RMSE are expressed in kcal/mol. Pearson and Spearman correlation coefficients and number of data points are also reported.}
\label{fig:preds_per_pdb_rosetta_both}
\centering
\end{figure}

Fig. \ref{fig:preds_per_pdb_rosetta_both} shows the predictions generated by pre-trained eGRAL-noESM (top row) and eGRAL-ESM (bottom row) on ROSETTA\textsubscript{sim,test} for 4 different PDB IDs: the plots report also the Pearson and Spearman correlation coefficients and the number of data points. The 4 PDBs are chosen to be a limited representative sample of performance on both ROSETTA\textsubscript{sim,test} and SKEMPI\textsubscript{cl,test}.

%%%%%%%%%%%%%%%%%%%%%%%%%%%%%%%%%%%%%%%%%%%%%%%%%%%%%%%%%%%%%%%%%%%%%%%%
%%%%%%%%%%%%%%%%%%%%%%%%%%%%%%%%%%%%%%%%%%%%%%%%%%%%%%%%%%%%%%%%%%%%%%%%
\subsection{SKEMPI\textsubscript{cl,test} add-ons}\label{app:skempi}
\begin{figure}[h]
\centering
\includegraphics[width=0.95\linewidth]{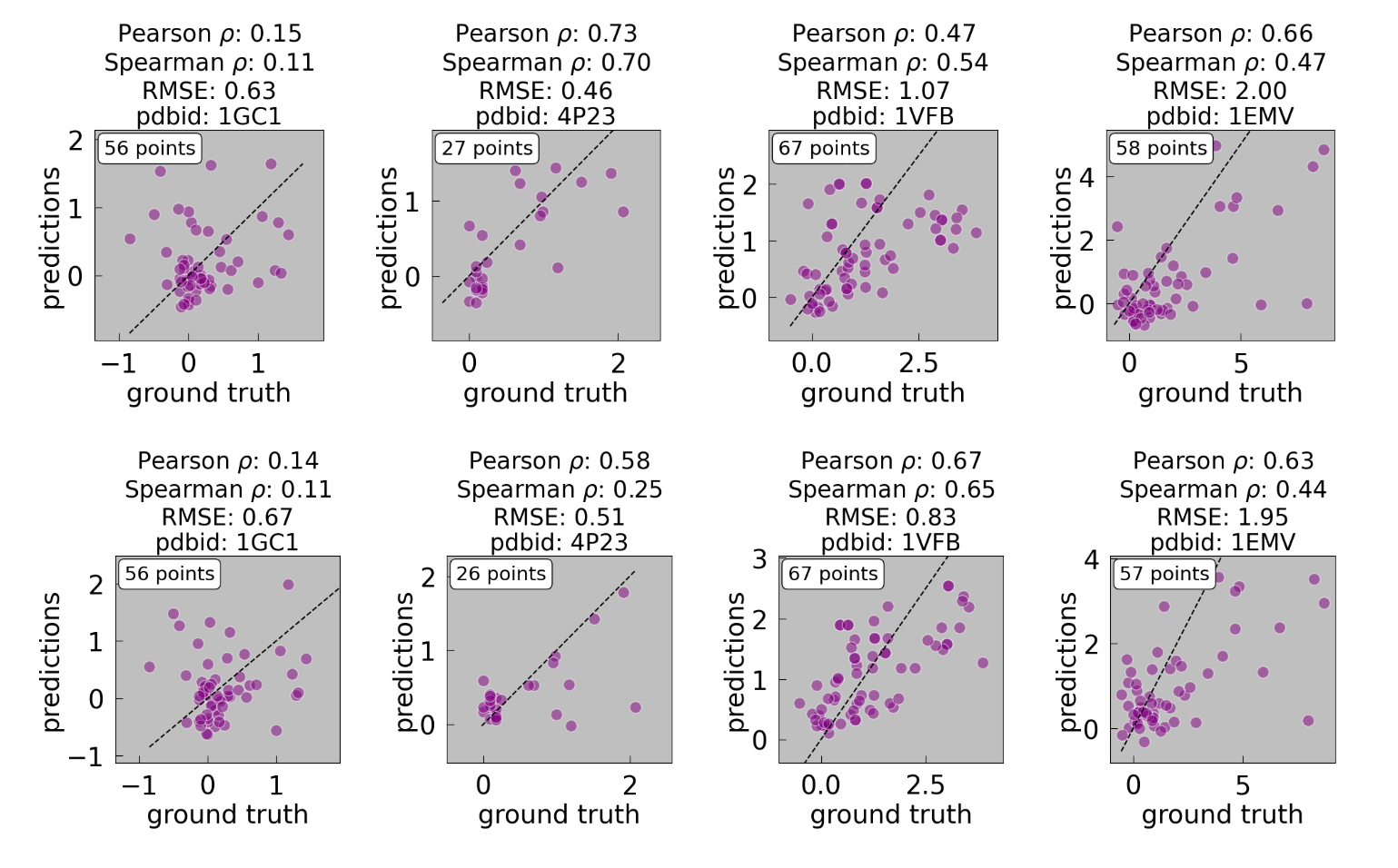}
\caption{Predicted results against ground truth by fine-tuned eGRAL-noESM (top row) and eGRAL-ESM (bottom row) on SKEMPI\textsubscript{cl,test}: the predictions are reported for 4 PDB IDs. Scores and RMSE are expressed in kcal/mol. Pearson and Spearman correlation coefficients and number of data points are also reported.}
\label{fig:preds_per_pdb_skempi_both}
\centering
\end{figure}

Fig. \ref{fig:preds_per_pdb_skempi_both} shows the predictions generated by fine-tuned eGRAL-noESM (top row) and eGRAL-ESM (bottom row) on SKEMPI\textsubscript{cl,test} for 4 different PDB IDs: the plots report also the Pearson and Spearman correlation coefficients and the number of data points. The 4 PDBs are chosen to be as less biased as possible towards too poor or too good performance on both ROSETTA\textsubscript{sim,test} and SKEMPI\textsubscript{cl,test}.

%%%%%%%%%%%%%%%%%%%%%%%%%%%%%%%%%%%%%%%%%%%%%%%%%%%%%%%%%%%%%%%%%%%%%%%%
%%%%%%%%%%%%%%%%%%%%%%%%%%%%%%%%%%%%%%%%%%%%%%%%%%%%%%%%%%%%%%%%%%%%%%%%
\subsection{RBD dataset}\label{app:rbd_test}
\begin{figure}[h]
\centering
\includegraphics[width=0.9\linewidth]{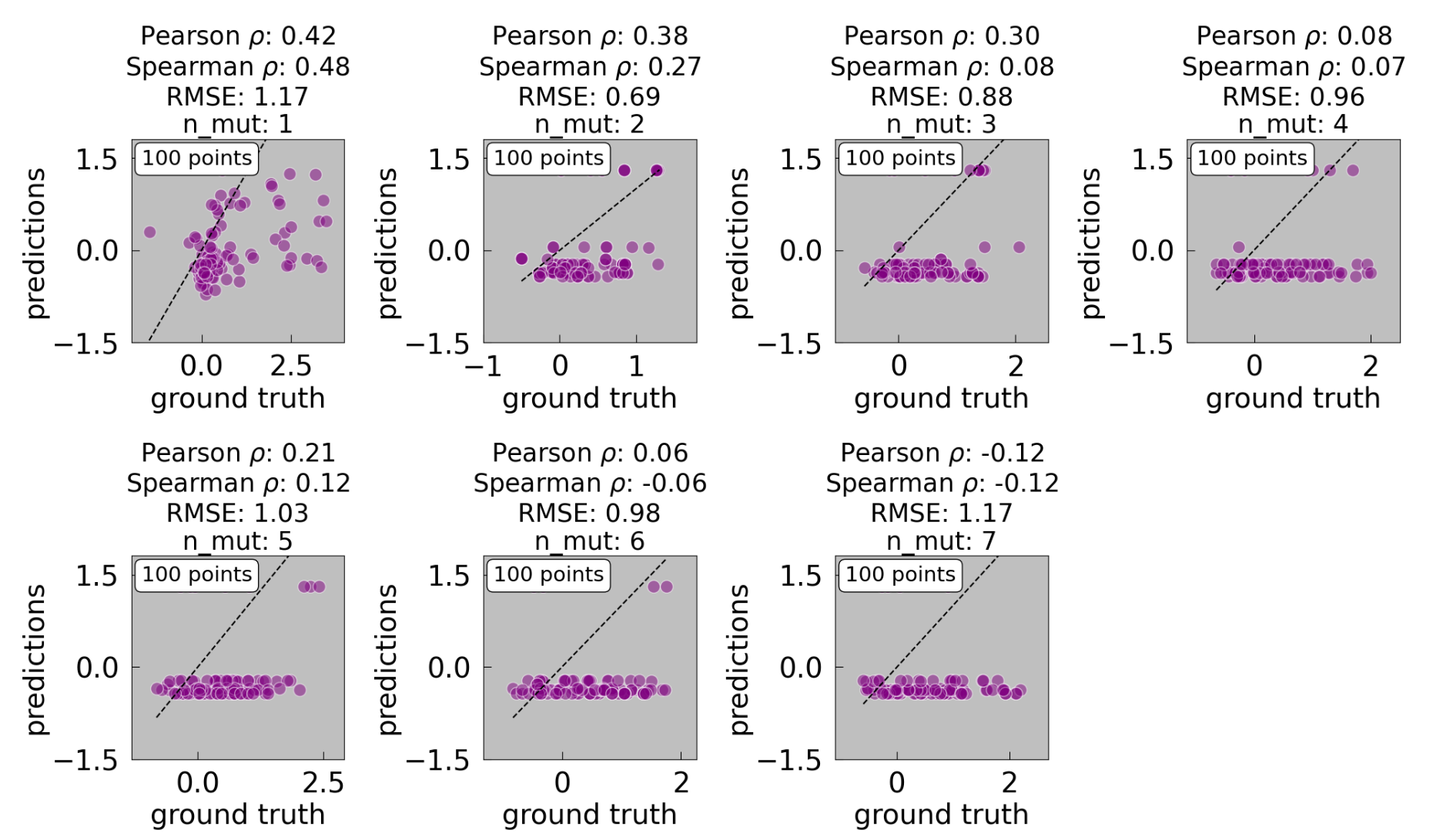}
\caption{Predicted results against ground truth by fine-tuned eGRAL-noESM on RBD\textsubscript{test}: the predictions are reported per number of mutations. Scores and RMSE are expressed in kcal/mol. Pearson and Spearman correlation coefficients and number of data points are also reported.}
\label{fig:preds_per_mut_rbd_noesm}
\centering
\end{figure}

\begin{figure}[h]
\centering
\includegraphics[width=0.9\linewidth]{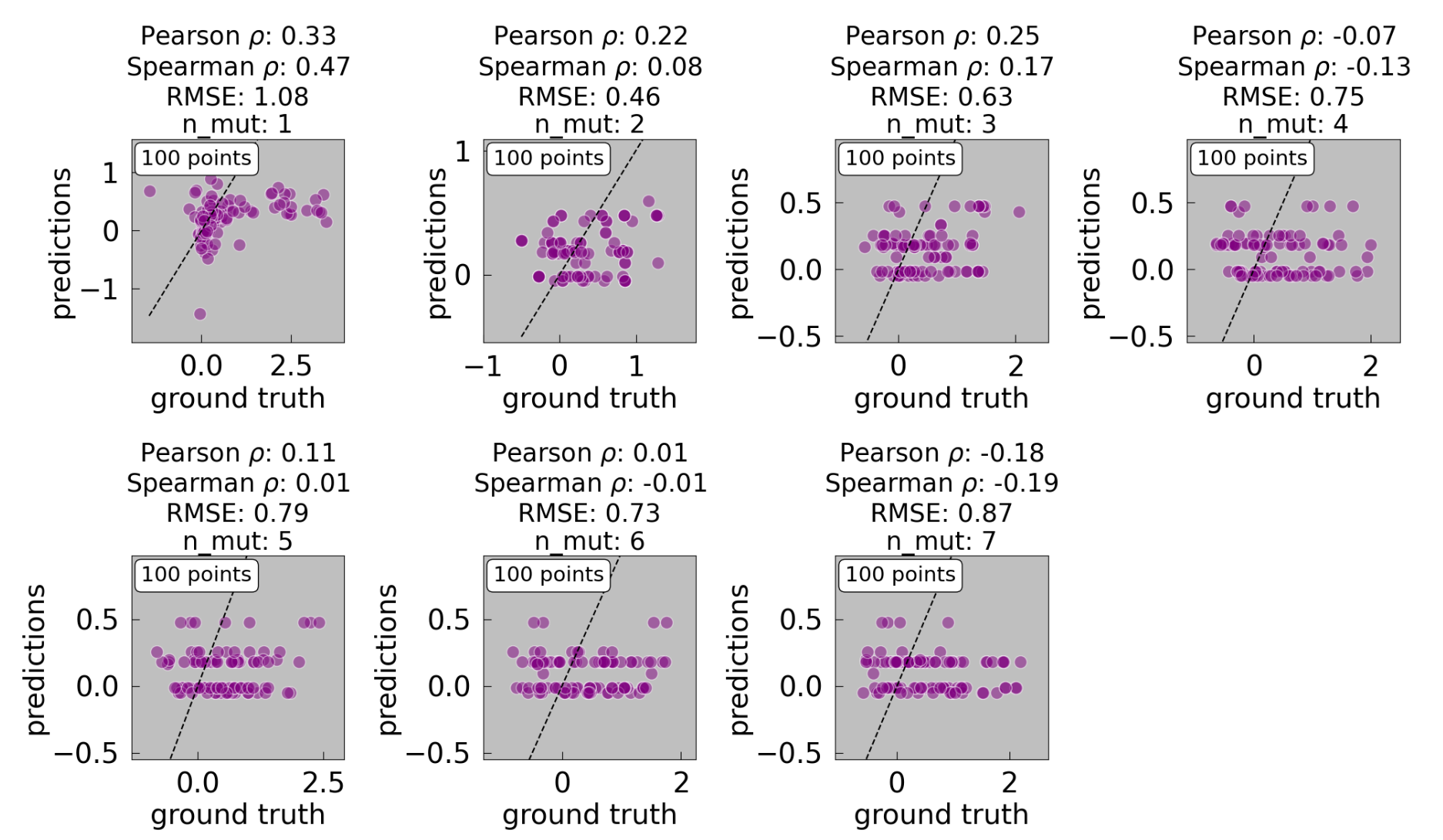}
\caption{Predicted results against ground truth by fine-tuned eGRAL-ESM on RBD\textsubscript{test}: the predictions are reported per number of mutations. Scores and RMSE are expressed in kcal/mol. Pearson and Spearman correlation coefficients and number of data points are also reported.}
\label{fig:preds_per_nmut_rbd_esm8m}
\centering
\end{figure}

Fig. \ref{fig:preds_per_mut_rbd_noesm} and \ref{fig:preds_per_nmut_rbd_esm8m} show the predictions generated on RBD\textsubscript{test} by fine-tuned eGRAL-noESM and eGRAL-ESM, respectively. The plots report also the Pearson and Spearman correlation coefficients and the number of data points. It is apparent that both models can output meaningful predictions for RBD\textsubscript{test} only in case of single point mutations.

%%%%%%%%%%%%%%%%%%%%%%%%%%%%%%%%%%%%%%%%%%%%%%%%%%%%%%%%%%%%%%%%%%%%%%%%
%%%%%%%%%%%%%%%%%%%%%%%%%%%%%%%%%%%%%%%%%%%%%%%%%%%%%%%%%%%%%%%%%%%%%%%%
\subsection{Hyper-parameters}\label{app:hparams}
This sections presents the hyper-parameters of eGRAL. Table \ref{tab:architecture} shows the architecture details of the scorer for both eGRAL-noESM and eGRAL-ESM. Table \ref{tab:h-params} instead shows the hyper-paramters used to build the graphs, and pre-train and fine-tune the models. Whereas for the fine-tuning phase we expected to need to decrease the learning rate and increase the dropout rate, given the small size of SKEMPI\textsubscript{cl,train}, we found the best results with the same parameters using during the pre-training.

\renewcommand{\arraystretch}{1.1}
\begin{table*}[h]
\caption{Architecture details of eGRAL scorer for both models: with and without nodes featurized with ESM. Following typical Haiku implementations, only the output sizes of the different linear layers within the used MLP are presented. The naming of the modules from the EGCL layers follows \citep{satoras2021} naming. Other naming follows \citep{egnn}.}
\label{tab:architecture}
\vskip 0.15in
\begin{center}
\setlength{\tabcolsep}{3pt} 
\begin{scriptsize}
\begin{sc}
\begin{tabular}{l|cc|cc|c}
\toprule
\multirow{2}{*}{} & \multicolumn{2}{|c}{Layer(s) size} & \multicolumn{2}{|c|}{Activation function} & \multirow{2}{*}{}\\
\cline{2-5}
 & eGRAL-noESM & eGRAL-ESM & eGRAL-noESM & eGRAL-ESM\\ 
\midrule
$\Phi_e$ & [6,8] & [16, 32] & Swish & Swish & \multicolumn{1}{c}{\multirow{3}{*}{\begin{tabular}[c]{@{}c@{}}2 EGCL layers\\ Haiku implemented\\ MLP net\end{tabular}}} \\
$\Phi_x$ & [8, 1] & [8, 1] & Swish & Swish \\
$\Phi_h$ & [6, 8] & [16, 32] & Swish & Swish \\
\cline{1-6}
Node features embedder & [8] & [256, 32] & None & None & \multicolumn{1}{c}{\multirow{3}{*}{\begin{tabular}[c]{@{}c@{}}Haiku implemented\\ MLP net\end{tabular}}} \\
Edge features embedder & [8] & [8] & None & None \\
Pre scoring & [10] & [10] & None & None \\
\cline{1-6}
Output & [1] & [1] & None & None  & \multicolumn{1}{c}{\begin{tabular}[c]{@{}c@{}}Haiku implemented\\ linear layer\end{tabular}} \\
\bottomrule
\end{tabular}
\end{sc}
\end{scriptsize}
\end{center}
\vskip -0.1in
\end{table*}

\renewcommand{\arraystretch}{1.1}
\begin{table*}[h]
\caption{Hyper-parameters used for graph building, pre-training and finetuning of eGRAL-noESM and eGRAL-ESM.}
\label{tab:h-params}
\vskip 0.15in
\begin{center}
\setlength{\tabcolsep}{3pt} 
\begin{scriptsize}
\begin{sc}
\begin{tabular}{l|cc|cc}
\toprule
\multirow{2}{*}{} & \multicolumn{2}{|c}{Embedder} & \multicolumn{2}{|c}{Scorer}\\
\cline{2-5}
 & eGRAL-noESM & eGRAL-ESM & eGRAL-noESM & eGRAL-ESM\\ 
\midrule
Learning rate & $3e^{-4}$ & $3e^{-4}$ & $3e^{-4}$ & $3e^{-3}$\\
Weight decay (AdamW) & $1e^{-4}$ & $1e^{-4}$ & $2e^{-2}$ & $2e^{-2}$\\
Dropout rate & - & - & $2e^{-2}$ & $3e^{-2}$\\
Batch size & $96$ & $96$ & $96$ & $96$ \\
Max number of nodes & $500$ & $500$ & $80$ & $80$ \\
Max number of nodes per batch & $500\cdot$BatchSize & $500\cdot$BatchSize & $80\cdot$BatchSize & $80\cdot$BatchSize \\
Max number of edges per batch & $500^{1.5}\cdot$BatchSize & $500^{1.5}\cdot$BatchSize & $80^{1.5}\cdot$BatchSize & $80^{1.5}\cdot$BatchSize \\
\bottomrule
\end{tabular}
\end{sc}
\end{scriptsize}
\end{center}
\vskip -0.1in
\end{table*}

\subsection{PDB cleaning}\label{app:pdb_clean_up}
We used a combination of pdbcleaner and openmm to fill missing hydrogens and relax them. Protein structures preparation was carried out using a combination of OpenMM and PDBFixer  \cite{eastman2017openmm} to rectify common issues found in Protein Data Bank (PDB) files, such as missing residues and nonstandard atoms. First crystallization artifacts such as additional solvents were removed, and missing residues were added using PDBFixer to ensure structural integrity. Structures were protonated, at pH 7 using the "amberfb15" force field. A thousand minimization steps using FIREMinimizer were then performed to relax the structure. Finally, atoms were renamed to conform to Rosetta \citep{das2008macromolecular} atoms naming conventions for further mutations and scoring.

\subsection{Variance analysis}\label{app:variance}
In this appendix we provide a variance analysis for the two models, eGRAL-noESM and eGRAL-ESM, trained on 5 different initialization seeds and 5 different data splits; architecture and hyperparameters for the pre-training and fine-tuning stages are the same as indicated in Appendix \ref{app:hparams}. Specifically, Table \ref{tab:seed_analysis} shows the performance on SKEMPI\textsubscript{cl,test} of the two models pre-trained and fine-tuned with the same splits of Appendix \ref{app:data_split} but with different initialization seeds (the first seed corresponding to 42 is the one adopted for the results presented in the main body of the paper). Conversely, Table \ref{tab:split_analysis} shows the performance of the two models pre-trained and fine-tuned with a initialization seed of 42 but 5 different training and validation splits (the first split is what used in the main body of the paper and explained in Appendix \ref{app:data_split}, whereas the test split always is not modified). For the sake of completeness, listed here are the list of PDB IDs of the 5 validation splits adopted (the training splits can be derived from the remaining PDB IDs of the totality of training and validation split of Appendix \ref{app:data_split}):
\begin{itemize}
    \item split 1: \small{1N80, 1TM1, 2HLE, 2OOB, 2QJA, 2VLQ, 3KBH, 3LZF, 3N0P, 3SZK, 4FZA, 4MNQ, 4OFY, 4UYQ, 5TAR, 5UFE.}
    \item split 2: \small{1B41, 1BJ1, 1JTG, 1SGN, 1XGP, 2B0U, 2BNR, 2JEL, 2VIR, 3KBH, 3N06, 3Q3J, 4EKD, 4J2L, 4YFD, 5UFE.}
    \item split 3: \small{1A22, 1GUA, 1KAC, 1MAH, 1MHP, 1MQ8, 1Y1K, 2B42, 2CCL, 3C60, 3HFM, 3SE3, 4GU0, 4JGH, 4Y61, 5TAR.}
    \item split 4: \small{1CT0, 1FSS, 1KTZ, 1N8Z, 1REW, 1TM1, 1TM7, 1XGR, 1Z7X, 2GOX, 2OI9, 3BT1, 3N85, 3NPS, 3SGB, 4G2V.}
    \item split 5: \small{1OGA, 1PPF, 1SGP, 1TMG, 1XGU, 2BTF, 2OOB, 2REX, 2VLO, 2WPT, 3L5X, 3Q8D, 3RF3, 4E6K, 4KRP, 4UYQ.}
\end{itemize}

\renewcommand{\arraystretch}{1.1}
\begin{table*}[h]
\caption{Evaluation metric summary for the fine tuned model on SKEMPI\textsubscript{cl,test} for 5 different seeds. RMSE is expressed in kcal/mol, $\rho_p$ is the Pearson correlation coefficient and $\rho_s$ is the Spearman rank correlation coefficient. The average metrics $\pm$ standard deviation across the seeds result RMSE=1.87$\pm$0.14, $\rho_p$=0.37$\pm$0.11, $\rho_s$=0.38$\pm$0.11 for eGRAL-noESM, RMSE=1.75$\pm$0.03, $\rho_p$=0.47$\pm$0.04, $\rho_s$=0.40$\pm$0.04 for eGRAL-ESM.}
\label{tab:seed_analysis}
\vskip 0.1in
\begin{center}
\setlength{\tabcolsep}{3pt} 
\begin{scriptsize}
\begin{sc}
\begin{tabular}{l|ccc|ccc|ccc|ccc|ccc}
\toprule
\multirow{2}{*}{} & \multicolumn{3}{|c}{SEED=42} & \multicolumn{3}{|c}{SEED=43} & \multicolumn{3}{|c}{SEED=44} & \multicolumn{3}{|c}{SEED=45} & \multicolumn{3}{|c}{SEED=46}\\
\cline{2-16}
 & RMSE & $\rho_p$ & $\rho_s$ & RMSE & $\rho_p$ & $\rho_s$ & RMSE & $\rho_p$ & $\rho_s$ & RMSE & $\rho_p$ & $\rho_s$ & RMSE & $\rho_p$ & $\rho_s$\\ 
\midrule
$eGRAL_{no ESM}^{fine-tuned}$ & 1.77 & 0.47 & 0.42 & 1.98 & 0.27 & 0.32 & 1.85 & 0.35 & 0.35 & \textbf{1.68} & \textbf{0.51} & \textbf{0.44} & 2.05 & 0.24 & 0.35\\
$Rosetta$ & 1.73 & 0.50 & 0.42 & 1.77 & 0.49 & 0.39 & 1.73 & 0.48 & 0.37 & 1.80 & 0.39 & 0.41 & 1.74 & 0.47 & 0.42\\
\bottomrule
\end{tabular}
\end{sc}
\end{scriptsize}
\end{center}
\vskip -0.15in
\end{table*}

\begin{figure}[h]
\centering
\includegraphics[width=0.6\linewidth]{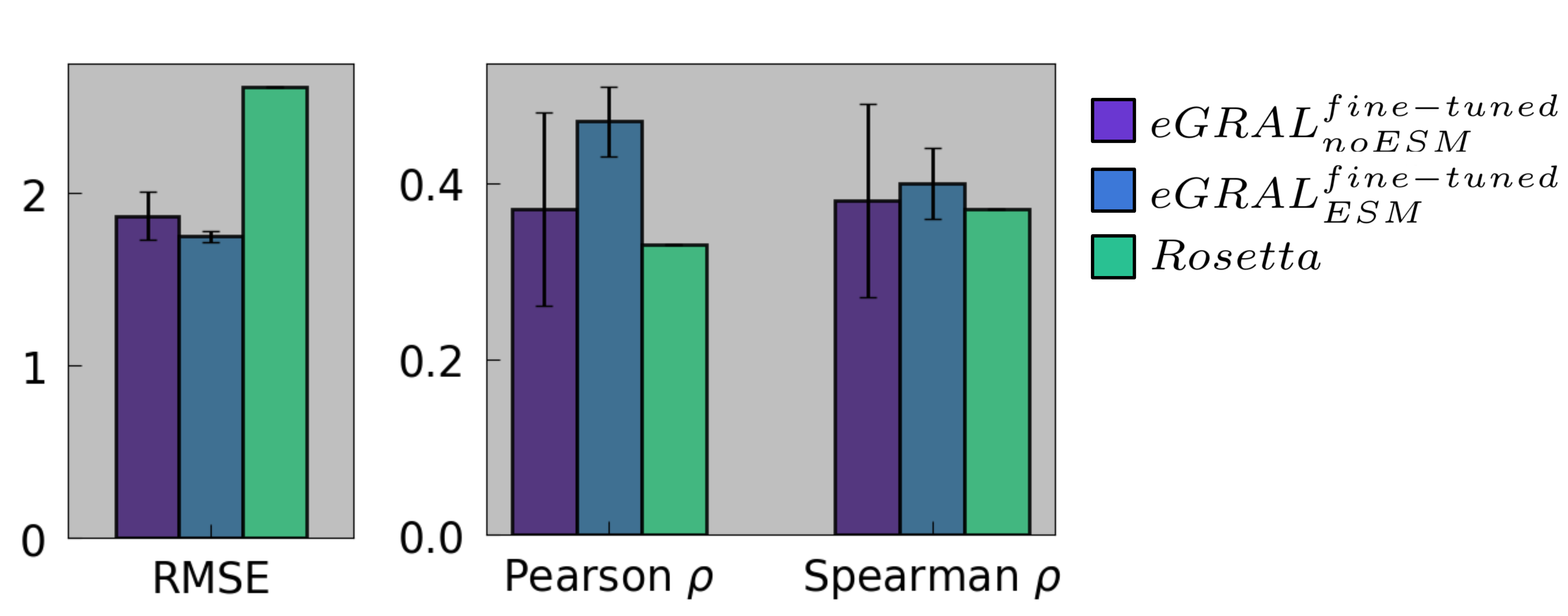}
\caption{Comparison of the performance of eGRAL-noESM (purple bar), eGRAL-ESM (blue bar) and Rosetta (green bar) for different initialization seeds. The performance of eGRAL-noESM and eGRAL-ESM are the average $\pm$ standard deviation of the results of Table \ref{tab:seed_analysis}. RMSE is expressed in kcal/mol.}
\label{fig:variance_analysis_seeds}
\centering
\end{figure}

\renewcommand{\arraystretch}{1.1}
\begin{table*}[h]
\caption{Evaluation metric summary for the fine tuned model on SKEMPI\textsubscript{cl,test} for 5 different splits. RMSE is expressed in kcal/mol, $\rho_p$ is the Pearson correlation coefficient and $\rho_s$ is the Spearman rank correlation coefficient. The average metrics $\pm$ standard deviation across the splits result RMSE=1.81$\pm$0.03, $\rho_p$=0.43$\pm$0.05, $\rho_s$=0.40$\pm$0.05 for eGRAL-noESM, RMSE=1.80$\pm$0.13, $\rho_p$=0.45$\pm$0.13, $\rho_s$=0.37$\pm$0.13 for eGRAL-ESM.}
\label{tab:split_analysis}
\vskip 0.1in
\begin{center}
\setlength{\tabcolsep}{3pt} 
\begin{scriptsize}
\begin{sc}
\begin{tabular}{l|ccc|ccc|ccc|ccc|ccc}
\toprule
\multirow{2}{*}{} & \multicolumn{3}{|c}{SPLIT 1} & \multicolumn{3}{|c}{SPLIT 2} & \multicolumn{3}{|c}{SPLIT 3} & \multicolumn{3}{|c}{SPLIT 4} & \multicolumn{3}{|c}{SPLIT 5}\\
\cline{2-16}
 & RMSE & $\rho_p$ & $\rho_s$ & RMSE & $\rho_p$ & $\rho_s$ & RMSE & $\rho_p$ & $\rho_s$ & RMSE & $\rho_p$ & $\rho_s$ & RMSE & $\rho_p$ & $\rho_s$\\ 
\midrule
$eGRAL_{no ESM}^{fine-tuned}$ & 1.77 & 0.47 & 0.42 & 1.80 & 0.43 & 0.42 & 1.86 & 0.35 & 0.41 & 1.79 & 0.47 & 0.39 & 1.83 & 0.41 & 0.36\\
$eGRAL_{ESM}^{fine-tuned}$ & 1.73 & 0.50 & 0.42 & \textbf{1.73} & \textbf{0.52} & \textbf{0.42} & 2.06 & 0.19 & 0.19 & 1.74 & 0.51 & 0.40 & 1.74 & 0.52 & 0.40\\
\bottomrule
\end{tabular}
\end{sc}
\end{scriptsize}
\end{center}
\vskip -0.15in
\end{table*}

\begin{figure}[h]
\centering
\includegraphics[width=0.6\linewidth]{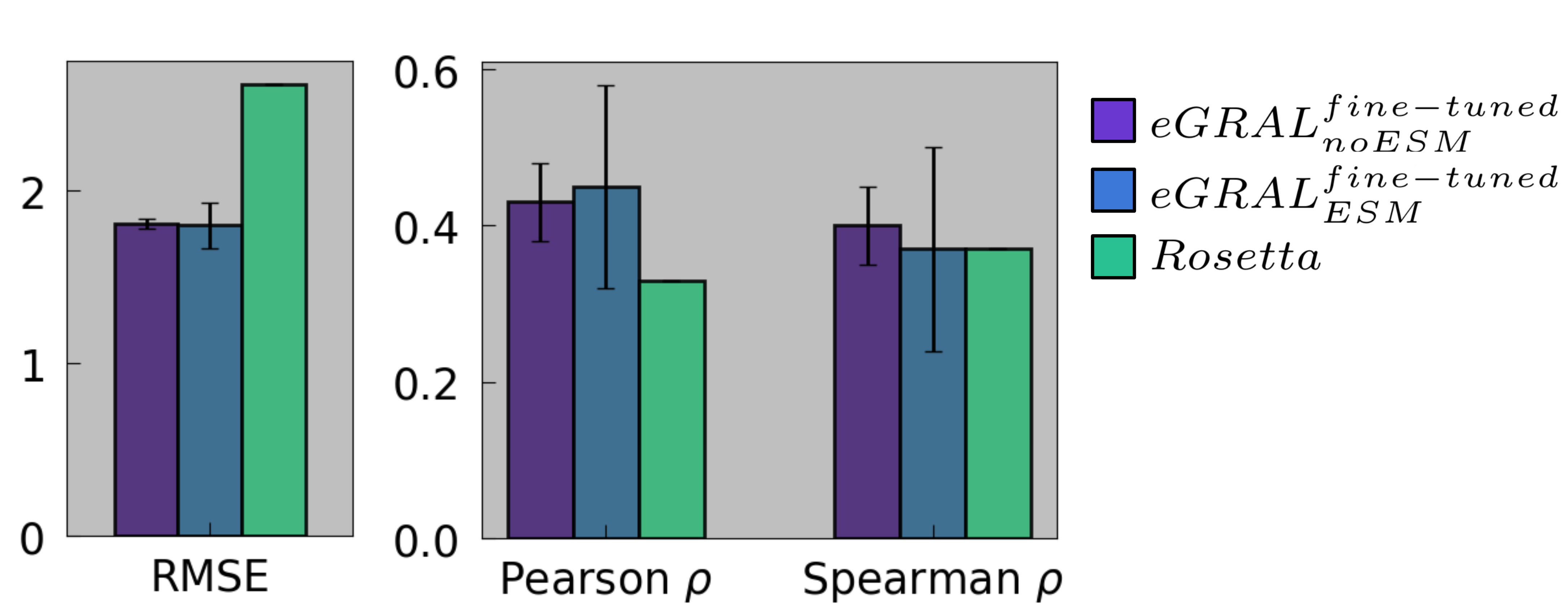}
\caption{Comparison of the performance of eGRAL-noESM (purple bar), eGRAL-ESM (blue bar) and Rosetta (green bar) for different data sees. The performance of eGRAL-noESM and eGRAL-ESM are the average $\pm$ standard deviation of the results of Table \ref{tab:split_analysis}. RMSE is expressed in kcal/mol.}
\label{fig:variance_analysis_splits}
\centering
\end{figure}

\subsection{Rosetta Vs eGRAL: execution speed comparison}\label{app:time_benchmark}
In this appendix, we provide a comparison of the execution speed between eGRAL and Rosetta \citep{das2008macromolecular} when scoring the $\Delta\Delta G$ of single point mutations. Since both eGRAL and Rosetta adopt parallelization to score the substitutions, we compare the speed of the two models by scoring all the possible mutations of a residue. We choose PDB 1A22 (having a medium size between the PDB IDs reported in \ref{app:data_split}), and we compare all the mutations at position 160. Rosetta takes an average of 49 seconds to score each variant in parallel, whereas eGRAL-noESM takes 25 seconds and eGRAL-ESM takes 34 seconds.

\subsection{geoPPI}\label{app:geoppi}
Following the tutorial on their github page, we ran geoPPI on the subset of our RBD\textsubscript{test} test set containing single point mutations.
\begin{figure}[h]
\centering
\includegraphics[width=0.35\linewidth]{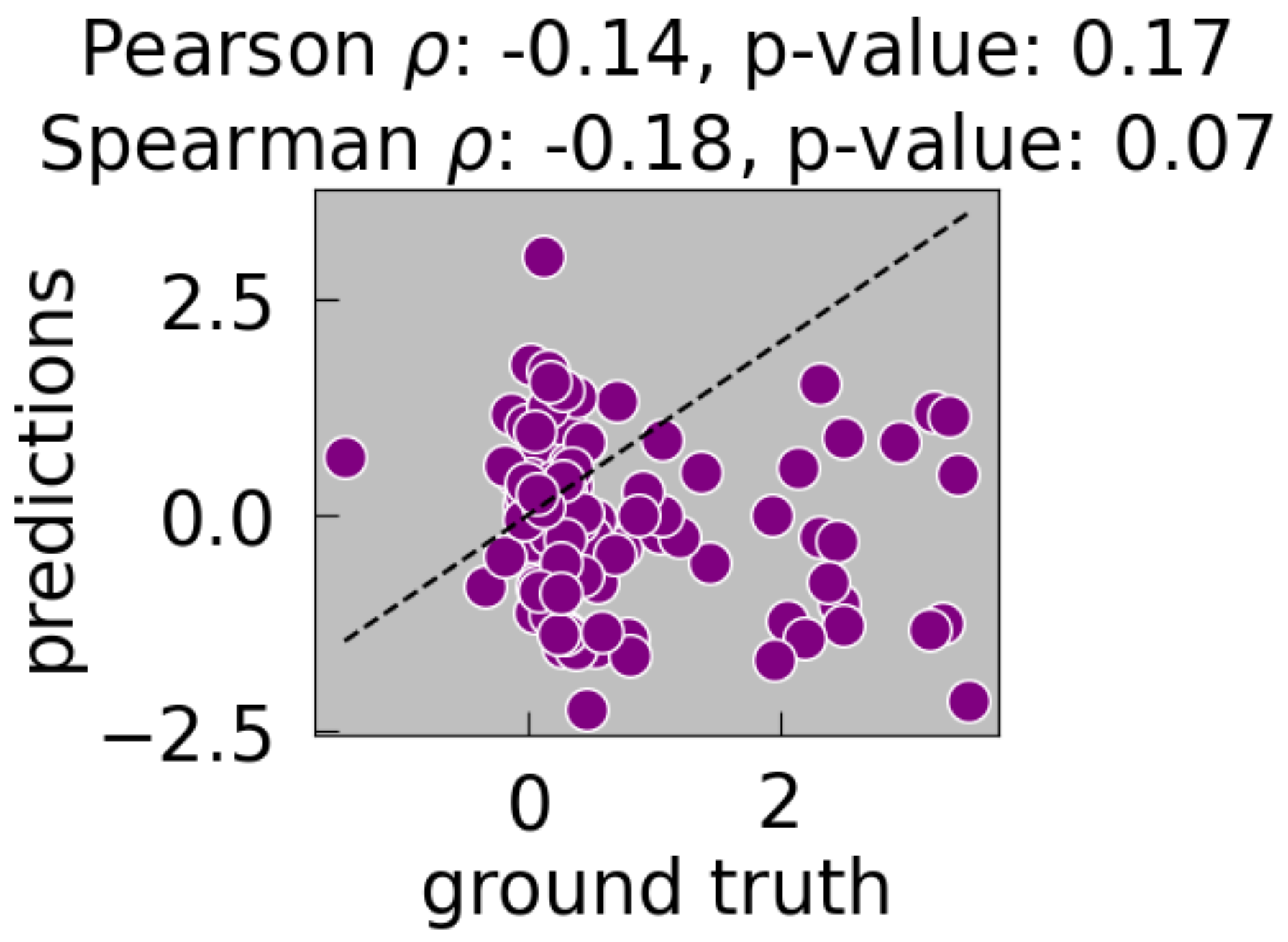}
\caption{Comparison between the ground truth of single point mutation $\Delta\Delta G$ in our RBD\textsubscript{test} dataset and predictions by geoPPI.}
\label{fig:geoPPI}
\centering
\end{figure}

\end{document}